%% file: main.tex
\documentclass[aps,prl,reprint,longbibliography,noeprint]{revtex4-2}
\usepackage{lipsum}

\usepackage{siunitx}
\usepackage{amsmath,bm}
\usepackage{amsbsy}
\usepackage{amssymb}
\usepackage{mathptmx, textcomp}
\usepackage{color}
\usepackage{braket}
\usepackage{graphicx}
\usepackage{textcomp}
\usepackage{notes2bib}
\usepackage{textcomp}
\usepackage{mwe}
\usepackage{siunitx}
\usepackage{filecontents}
\usepackage{soul}

\usepackage{svg}

\definecolor{bl}{rgb}{0, .1, .6}
\definecolor{gr}{rgb}{.2, .6, .2}
\usepackage[colorlinks=true, citecolor = bl, linkcolor = bl, urlcolor=bl, pdfborder={0 0 0}]{hyperref}

\newcommand{\change}[1]{#1}
\DeclareSIUnit\gauss{G}
\DeclareSIUnit\photon{ph}
\DeclareSIUnit\pixel{px}

\begin{document}
\title{Trapping and imaging single dysprosium atoms in optical tweezer arrays}

\author{Damien Bloch}\email{damien.bloch@institutoptique.fr}
\author{Britton Hofer}
\author{Sam R.~Cohen}\altaffiliation{Present address: Department of Physics, Stanford University, Stanford, California 94305, USA}
\author{Antoine Browaeys}
\author{Igor Ferrier-Barbut}\email{igor.ferrier-barbut@institutoptique.fr}
\affiliation{Universit\'e Paris-Saclay, Institut d'Optique Graduate School, CNRS, 
Laboratoire Charles Fabry, 91127, Palaiseau, France}

\begin{abstract}
We report the preparation and observation of single atoms of dysprosium in arrays of optical tweezers with a wavelength of $\SI{532}{\nano\meter}$, imaged on the intercombination line at $\SI{626}{\nano\meter}$. We use the anisotropic light shift specific to lanthanides and in particular a large difference in tensor and vector polarizabilities between the ground and excited states to tune the differential light shift and produce tweezers in near-magic or magic polarization. This allows us to find a regime where single atoms can be trapped and imaged. 
Using the tweezer array toolbox to manipulate lanthanides will open new research directions for quantum physics studies by taking advantage of their rich spectrum, large spin and magnetic dipole moment. 

\end{abstract}

\maketitle


Trapping and cooling of single atoms in tweezer arrays \cite{Schlosser2001,Nogrette2014} has allowed tremendous progress in quantum science and metrology \cite{Browaeys2020,Kaufman2021}. 
These techniques were first used on alkali atoms \cite{Barredo2016,Endres2016,Kim2016}, before being extended to alkaline-earth species \cite{Cooper2018,Norcia2018,Saskin2019} and molecules \cite{Anderegg2019}. 
In parallel to this progress, experiments with quantum gases of lanthanides have explored dipolar physics  \cite{Chomaz2023} and topology \cite{Burdick2016,Chalopin2020} among other examples. 
Controlling lanthanides in single-atom tweezers will offer new possibilities for exploiting their specific properties. 
Their anisotropic light-matter interaction \cite{Becher2018,Chalopin2018a} results in a broad tunability of trapping potentials useful to produce sub-wavelength interatomic distances \cite{Nascimbene2015,du2023} or for quantum-enhanced sensing \cite{Chalopin2018b}. 
Dimers with a large magnetic dipole moment \cite{Frisch2015,Maier2015} or atoms with an electric dipole \cite{Lepers2018, Kirilov2023} might be produced to study quantum magnetism \cite{dePaz2013} in tweezer arrays. 
Finally, their many transitions from the ground state, spanning a broad range of wavelengths and linewidths makes them an interesting platform for studies of collective light-matter interactions \cite{Bettles2016,Shahmoon2017,Rui2020}. 
In this Letter we demonstrate single-atom trapping of dysprosium in optical tweezers, imaging on the narrow intercombination line by making use of the strong anisotropic light shift of Dy.

The rich spectrum of optical transitions of lanthanides has been used to operate efficient laser cooling and produce degenerate quantum gases \cite{Chomaz2023}. 
Transitions from the $6s^2$ electrons are similar to those of two-electron atoms such as Yb and Sr, and the methods developed to prepare single atoms of these species can be adapted to lanthanides. 
Here we rely on the intercombination line between $G=4f^{10}6s^2\,^5I_8$ and $ E=4f^{10}(^5I_8)6s6p(^3P^\circ_1)\,(8,1)^\circ_9$ of Dy, generally used for magneto-optical traps \cite{Maier2014,Dreon2017}, to image single Dy atoms. 
This transition has a wavelength $\lambda = \SI{626}{\nano\meter}$ and a linewidth $\Gamma=2\pi\times\SI{135}{\kilo\hertz}$.
Another advantage of lanthanides is their non-vanishing vector and tensor polarizabilities. 
The tensor polarizability was recently used to demonstrate magic trapping for the Dy intercombination transition at a trap wavelength of $\SI{1070}{nm}$ \cite{Chalopin2018a}. 
We rely in this work both on the tensor and vector polarizabilities \cite{Kim2013} to obtain magic trapping at $\SI{532}{nm}$.

\begin{figure*}
    \centering
    \includegraphics{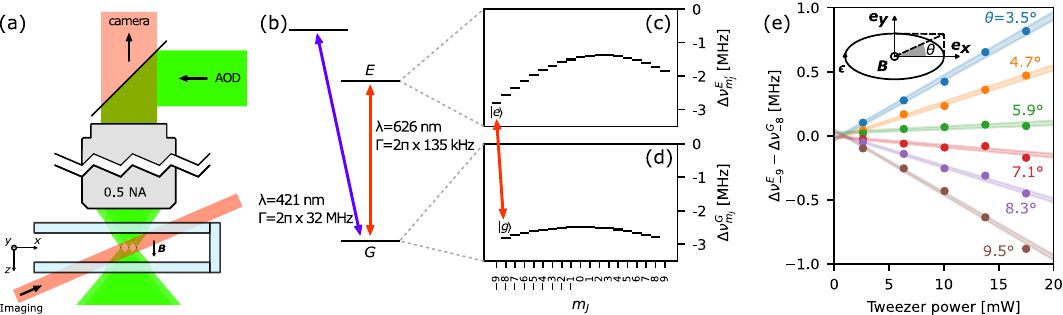}
    \caption{(a)~Simplified diagram of the beams used to trap and image single atoms in the tweezers. MOT beams are not shown here. 
    (b)~Relevant energy levels of Dy and their associated optical transitions used in this work. 
    (c)~Anisotropic light shift experienced by the Zeeman states in both the ground $\Delta\nu^G_{m_J}$ and (d)~excited $\Delta\nu^E_{m_{J'}}$ manifolds in magic conditions. 
    The values plotted here are obtained by diagonalizing the full hamiltonian including the Zeeman effect and the trap light shift. 
    We then subtract the Zeeman shift to the eigenvalues of the hamiltonian, to keep only the light shift.
    (e)~Measured frequency difference between $\ket g$ and $\ket e$ with respect to their unperturbed frequency as a function of tweezer power and for different ellipticities $\theta$ of the tweezer polarization, shaded areas are linear fits with confidence interval.}
    \label{fig:setup}
\end{figure*}

We generate $5\times5$ tweezer arrays with \SI{5}{\micro\meter} spacing at a wavelength of \SI{532}{\nano\meter} \footnote{The trapping laser is a Coherent Verdi V10 with measured wavelength \SI{532.208}{\nano\meter}.} using a 2D acousto-optic deflector (AOD) driven by a multitone signal \cite{Endres2016,SM}.
The tweezer light is sent through a 0.5-numerical aperture (NA) microscope objective (Mitutoyo G Plan Apo 50X) placed outside a glass cell, resulting in a tweezer waist $w_0\approx\SI{500}{\nano\meter}$ \footnote{The waist is defined as the $1/e^2$ radius of a gaussian beam that fits best the expected radial intensity profile.}.
Each trap has a power of \SI{2}{\milli\watt}, yielding a potential depth of about \SI{150}{\micro\kelvin}.
Our setup is schematized in Fig.\,\ref{fig:setup}(a) and more details will be published in \cite{Bloch2023}.
We use the $^{162}$Dy isotope in this work.
The experiment begins with a 2D magneto-optical trap (MOT) on the broad transition of Dy at \SI{421}{\nano\meter}, as in \cite{jin2023}, to cool and redirect atoms towards a glass cell.
In the glass cell, we capture the atoms with a two color core-shell MOT \cite{Lee2015} and eventually transfer them to a MOT using only the narrow intercombination line.
Following the MOT loading stage, the atoms are pumped in the lowest Zeeman state $\ket g=\ket {G, J=8, m_J=-8}$ by ramping the intensity to $I=0.1\, I_{\rm sat}$, with $I_{\rm sat}=\SI{72}{\micro\watt\per\centi\meter^2}$, and detuning to $\Delta=-(2\pi)\,\SI{1.5}{\mega\hertz}$ \cite{Maier2014,Dreon2017}.
The tweezers are overlapped for \SI{100}{\milli\second} on the MOT. After this, each trap is filled with more than one atom on average.


Dysprosium has a large Zeeman manifold in both the ground state ($J=8$) and excited state ($J'=9$).
This strongly influences imaging and cooling since the scattering rate on a narrow transition depends on the atom's internal state. 
We apply a magnetic field of \SI{7}{G} to isolate a closed $\sigma^-$ transition between $\ket g$ and $\ket e = \ket {E, J'=9, m_J'=-9}$.
This leaves the $\pi$ ($m_J=-8 \leftrightarrow m_J'=-8$) and $\sigma^+$ ($m_J=-8 \leftrightarrow m_J'=-7$) transitions strongly off-resonance, respectively detuned by about \SI{13}{\mega\hertz} and \SI{25}{\mega\hertz} ($~95\,\Gamma$ and $~190\,\Gamma$, resp.). It ensures negligible photon scattering rates for these transitions
and the atoms are then imaged solely on the cycling $\sigma^-$ transition.

To obtain single atoms we induce light-assisted collisions that eject pairs of atoms from the multiply-loaded tweezers \cite{Schlosser2001}. 
We observe that such collisions take place in a few milliseconds when shinning red-detuned light. The collision pulse lasts for \SI{10}{\milli\second} and has the same parameters as used for imaging specified below.
After this, the tweezers are randomly loaded with zero or one atom, with a filling fraction close to 50\%.

Next, to image single atoms, we need to precisely tune the trapping potential. 
Indeed for such a narrow linewidth,
high fidelity single-atom imaging requires magic trapping where $\ket g$ and $\ket e$ have the same polarizability \cite{Saskin2019,Ma2022}.
Whether or not such a condition exists for a given species depends in general on the trapping wavelength. 
In contrast with other species, the strong anisotropy of the polarizability of lanthanides allows one to tune the differential polarizability between $\ket g$ and $\ket e$ by changing the tweezer polarization \cite{Becher2018,Chalopin2018a}.
This can lead to magic trapping in broad ranges of wavelengths.
Measurements of the scalar, vector and tensor polarizabilities for both the ground ($G$) and excited ($E$) manifolds at \SI{532}{\nano\meter} will be reported in \cite{Bloch2023}.
We use the large vector polarizability of the excited state and create an elliptic polarization of the tweezers, with Jones vector $(\epsilon_x,\epsilon_y)=(\cos \theta,i\,\sin\theta)$ in the plane perpendicular to the magnetic field. Fig.\,\ref{fig:setup}(e) shows the shift of the transition measured with fluorescence spectroscopy as a function of trap power for different ellipticities $\theta$. We find an ellipticity $\theta\simeq +6^\circ$ for which the transition $\ket g \leftrightarrow \ket e$ is magic [see fig. \ref{fig:setup}(c, d)].

\begin{figure}[b]
    \centering    
    \includegraphics{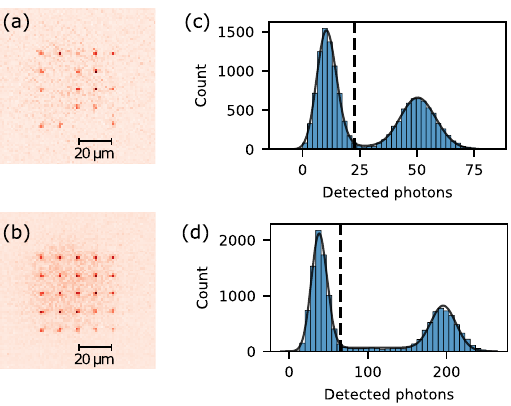}
    \caption{(a) Single shot and (b) average picture of $5\times5$ trap arrays for exposure time of \SI{30}{\milli\second}. The halo that can be seen between traps on the average image is due to the fluorescence of the microscope objective. (c) Histogram of the fluorescence of the central trap for \SI{30}{\milli\second} (d) and \SI{100}{\milli\second} exposure time. The line is a fit to the sum of two peak distributions joined by a ``bridge'' (see text). Dashed lines indicate the chosen threshold to maximize the imaging fidelity $F$.}
    \label{fig:histograms}
\end{figure}
This magic trapping condition allows us to image single atoms in the tweezers.
Fluorescence is induced by a single non retro-reflected beam with propagation axis having components along both the radial and axial directions of the tweezers \cite{Saskin2019}, which is necessary to cool efficiently while imaging.
This beam is red-detuned by $\Delta = - 1.0 \, \Gamma$ 
and has an intensity $I=0.8 \, I_{\rm sat}$. The duration of the imaging pulse is typically \SI{30}{\milli\second}.
The light scattered by the atoms is collected onto a CMOS camera (Hamamatsu C15550-20UP) through the same microscope objective used to focus the traps. 
For a single shot image as in Fig.\,\ref{fig:histograms}(a), we count the number of collected photons in a small circular area around each trap.
We repeat the experiment, reloading the MOT and the tweezers for every shot, and we record the histogram of the collected fluorescence as shown in Fig.\,\ref{fig:histograms}(c). 
The histograms exhibit two peaks characteristic of the single-atom regime: one peak corresponding to zero atoms and the other peak, with about 50 photons detected, corresponding to a single atom in the trap.

These histograms are shifted and broadened by background light. 
This light is due to the tweezers beam at \SI{532}{\nano\meter} going through the microscope and causing \change{the glass of the lenses inside the objective} to fluoresce at longer wavelengths, including the imaging wavelength of \SI{626}{\nano\meter}. 
To mitigate this effect, two angle-tunable dichroic filters, one short-pass and one long-pass (Semrock TSP01-628 and TLP01-628), are placed on the path before the camera to transmit only a narrow wavelength band around \SI{626}{\nano\meter}. 
This reduces the light reaching the camera to about 20 photons per pixel per second for \SI{50}{\milli\watt} of \SI{532}{\nano\meter} light going through the microscope.
This remaining background can be seen in figure \ref{fig:histograms}(b).

To determine the presence of a single atom in a given picture, we compare the number of photons collected to a given threshold. 
If the fluorescence is higher than the threshold, we label the trap as containing an atom, otherwise we label it as empty.
In the following, we characterize the fidelity and induced losses of our imaging. The fidelity represents the probability to correctly label the \emph{initial} presence of an atom in a trap. In addition, losses might be induced by the imaging sequence through which a ground state atom initially present in the trap is not detected in a subsequent imaging pulse. Both infidelity and imaging-induced losses will limit the ability to image and re-arrange large atomic arrays \cite{Barredo2016,Endres2016}.  

The experimental fluorescence histograms are well modeled as the sum of three distributions. 
The first peak is centered on the number of background photons $N_0$, with area the empty-trap probability $P_0\simeq \SI{50}{\percent}$.
A second peak represents events where an atom is present for the full duration of the imaging.
It is centered on $N_0 + N_1$ where $N_1$ is the number of photons scattered by the atom.
Its area is $P_1\times P_{\rm survival}$ where $P_1=1-P_0$ is the probability to have \emph{initially} one atom in the trap and $P_{\rm survival}$ is the probability that the atom survives imaging. 
The third contribution is a flat distribution that bridges the two peaks, visible in Fig.\,\ref{fig:histograms}(d), that corresponds to the events where atoms are lost while they are being imaged \cite{Cooper2018}. 
Its area is $P_1 \times P_{\rm loss}$, with $P_{\rm loss} = 1 - P_{\rm survival}$. 
We give more details on the exact form used to model the distributions in \cite{SM}. 
Adjusting this model to the observed histograms, we extract the parameters $N_0$, $N_1$, $P_0$, $P_{\rm loss}$ and estimate the best threshold to maximize the imaging fidelity $F$ (see \cite{SM}).
All quantities above depend in general on every imaging parameter such as exposure, imaging intensity and detuning, as well as tweezer power.
We optimized them to have the highest imaging fidelity.

\begin{figure}
    \centering    
    \includegraphics{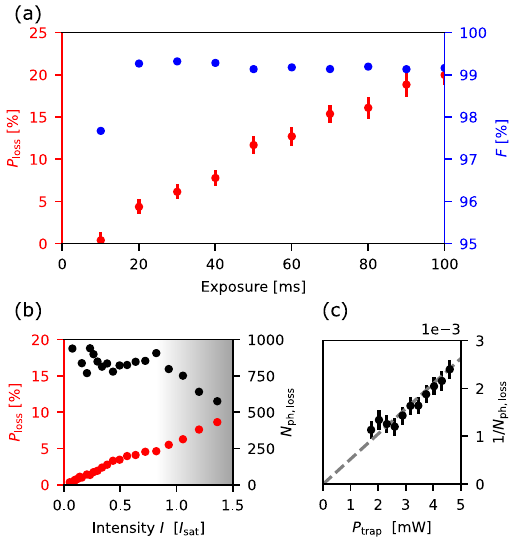}
    \caption{(a) Imaging fidelity and loss probability as a function of the exposure time. (b) Loss probability ($P_{\rm loss}$) and average number of \SI{626}{\nano\meter} photons detected before a loss ($N_{\rm ph,\,loss}$) as a function of imaging power for \SI{30}{\milli\second} exposure. In the shadowed area, losses are due to less efficient cooling. (c) \change{The inverse of $N_{\rm ph,\,loss}$ as a function of tweezer power. The dashed grey line is a linear fit.}}
    \label{fig:losses_vs_exposure}
\end{figure}

For example, we show in Fig.\,\ref{fig:losses_vs_exposure}(a) $F$ and $P_{\rm loss}$ for several exposure times.
At short duration, the fidelity is low because an atom does not scatter enough photons to be clearly distinguished from the background.
The fidelity increases with exposure, eventually reaching a maximum after a few tens of milliseconds.
However, the loss probability increases linearly with time. 
The imaging duration we choose is then a compromise between high fidelity and low losses.
In typical conditions, we image the atoms in \SI{30}{\milli\second}, which is resilient to small fluctuations of parameters and we reach $F=\SI{99.1(0.2)}{\percent}$ and $P_{\rm loss}=\SI{6.1(0.8)}{\percent}$ \footnote{This means that out of the 6\% of atoms that are lost during the imaging, most (roughly 5 in 6) are correctly labeled before being lost.}. 

To identify the origin of the losses, we measured the influence of the imaging parameters on $P_{\rm loss}$. 
We took a first picture to detect the atoms, then applied an imaging pulse for \SI{30}{\milli\second} varying the imaging parameters and finally measured the probability for the atom to have survived this pulse by taking a last image. The first and last pictures are taken with fixed parameters: \SI{30}{\milli\second}, $I=0.8\,I_{\rm sat}$, $\Delta=-\Gamma$. 
As shown in Fig.\,\ref{fig:losses_vs_exposure}(b), we observe that $P_{\rm loss}$ increases linearly with imaging power.
We also measure the average number of detected photons before an atom is lost $N_{\rm ph,\,loss} = -N_{\rm detected} / \ln(P_{\rm survival})$ \footnote{This comes from $P_{\rm survival}(t)=e^{-t/\tau_{\rm loss}}=e^{-N_{\rm detected}/N_{\rm ph,\,loss}}$}, where $N_{\rm detected}$ is the number of detected photons during the pulse.
For $I\lesssim I_{\rm sat}$, $N_{\rm ph,\,loss}$ is approximately constant. It decreases for higher intensities [gray area in Fig.\,\ref{fig:losses_vs_exposure}(b)] due to less efficient Doppler cooling \cite{Westbrook1989}. 
We also find that $N_{\rm ph,\,loss}$ is constant when varying the detuning for $\Delta \lesssim -1 \, \Gamma$.
Thus our observations suggest that the probability to lose an atom is directly proportional to the time it spends in the excited state $\ket e$.

This could be caused by a decay from $\ket e$ to dark or non-trapped states.
However, the intercombination transition is closed and we have also checked that the atoms are not pumped to other Zeeman states of the ground manifold. 
These losses are thus likely due to further excitation by the trapping light from $\ket e$ to a highly excited state in Dy's dense spectrum. 
We indeed observe that the leakage to non-imaged states increases with trap power: \change{Fig.\,\ref{fig:losses_vs_exposure}(c) shows the inverse of $N_{\rm ph,\,loss}$ for fixed imaging parameters as a function of trap power at \SI{532}{\nano\meter}. A linear increase is observed showing that a deeper trap means a higher loss probability per imaging photon.}
We thus conclude that losses are due to a two-photon event: an atom in $\ket e$ absorbs a trap photon, sending it to a highly excited state from which it then decays to non-imaged states. 
There indeed exists a state with a dipole-allowed transition with $\ket e$ ($4f^{10}5d6p$, $J=10$ at \SI{34776.04}{\per\centi\meter}) lying only about \SI{400}{\giga\hertz} away from the sum of the two laser frequencies \cite{Martin1978}. 
These losses are the main factor limiting imaging fidelity, and using a tunable trapping laser to increase the detuning from this state should allow to mitigate them. 
We expect this to be necessary for other lanthanides because of their dense spectrum.
\begin{figure}
    \centering
    \includegraphics{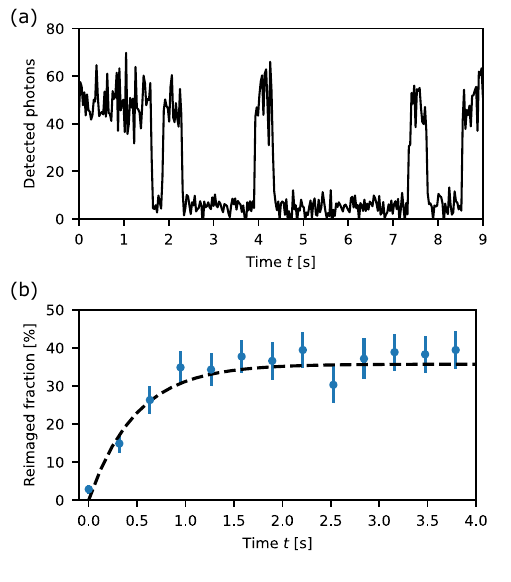}
    \caption{(a) Number of collected photons over \SI{30}{\milli\second} for a given trap under continuous illumination but with no background gas to reload the trap. This shows events where the atom is pumped to metastable states and events where it decays back to the ground state. (b) Probability to re-image an atom that previously became dark after having applied an imaging pulse of \SI{1.5}{\second}. The dashed line is a fit to an exponential saturation, with decay time \SI{0.48(0.08)}{\second}.}
    \label{fig:metastable}
\end{figure}

We further observe that dark atoms can decay back to $\ket g$ from metastable states. Indeed, a trap initially containing an atom and that became dark sometimes spontaneously becomes bright again although the MOT is turned off. 
This can be seen on Fig.\,\ref{fig:metastable}(a) where we plot the fluorescence of a single trap continuously imaged and observe discrete jumps from bright to dark and vice-versa. 
Starting from initially empty traps, we do not observe the appearance of atoms, ruling out reloading from residual background pressure. 
Similar observations were reported with Yb in \cite{Saskin2019}, identified as the excitation of the atom to metastable states and spontaneous decay to the ground state.
To measure the average time it takes for the atoms to come back, we apply a pulse of imaging light for \SI{1.5}{\second}. 
After this pulse, about 70\,\% of the atoms are no longer imaged.
We plot in Fig.\,\ref{fig:metastable}(b) the fraction of these dark atoms that subsequently reappear as a function of the wait time. 
We thus observe that \SI{35}{\percent} of them come back after a typical time $\tau=\SI{0.48(0.08)}{\second}$. From these measurements we extract a branching ratio of about 65\,\% of decay towards trapped metastable states versus non-trapped ones \cite{SM}. We leave for future research the exact identification of these states.

We finally measured the temperature and lifetime of atoms in the tweezers. The lifetime in particular is important in views of sorting atoms to form large ordered arrays \cite{Schymik2022}.
For this, we used the release and recapture method, see \cite{SM}.
Directly after imaging, we measured a temperature of $\SI{6.3(0.2)}{\micro\kelvin}$, slightly higher than the Doppler temperature for the intercombination transition ($T_D= \SI{3.2}{\micro\kelvin}$). Next, in shallow tweezers (depth $U_0=\SI{150}{\micro\kelvin}$, $P_{\rm trap} = \SI{2}{\milli\watt}$), we observed a heating rate of $\SI{1.7(0.2)}{\micro\kelvin\per\second}$, that limits the lifetime in the absence of cooling to about \SI{10}{\second}.
This heating rate is compatible with the off-resonant scattering of trap photons in the ground state.
Indeed from the calculated imaginary part of the polarizability at \SI{532}{\nano\meter} \cite{Li2017}, we expect a heating rate of a few microkelvins per second. We mitigated this heating by applying cooling light (intensity $I=5\times\,10^{-3}\, I_{\rm sat}$, detuning $\Delta=-1.3\, \Gamma$), and observed a lifetime of \SI{300(30)}{\second}, limited by the two-photon losses studied above (see \cite{SM}). 

In conclusion, we have demonstrated single-atom trapping and high-fidelity imaging of Dy on the intercombination line in tweezers rendered magic by fine tuning the tweezer polarization. 
Single-atom trapping of lanthanides opens exciting opportunities. 
For instance it can be used to obtain subwavelength distances using the anisotropic polarizability \cite{du2023} or also by directly loading an accordion lattice. 
This could be used to create atomic waveguides \cite{Asenjo2017}, or to prepare directly extended Bose-Hubbard models \cite{su2023} from optical tweezers.

\begin{acknowledgments}

We acknowledge fruitful discussions with Maxence Lepers and Jeff Thompson, experimental assistance by Florence Nogrette and critical reading of the manuscript by Thierry Lahaye and Giovanni Ferioli.  We note that another setup based on a similar 421-nm 2D-MOT loading a 626-nm 3D-MOT of Dy atoms has been developed in the group of L. Chomaz \cite{jin2023}. We have widely benefited from exchanges between our groups.
This project has received funding by the Agence Nationale de la Recherche (JCJC grant DEAR, ANR-22-PETQ-0004 France 2030, project QuBitAF) and by the European Union (ERC StG CORSAIR,  101039361, ERC AdG ATARAXIA 101018511).

\end{acknowledgments}

\bibliography{biblio1}

\clearpage
\clearpage

\begin{center}
	{\Large Supplemental Material}
\end{center} 
\setcounter{figure}{0}
\renewcommand\thefigure{S\arabic{figure}} 
\setcounter{equation}{0}
\renewcommand\theequation{S\arabic{equation}} 
\appendix

\section{Trap homogeneity}\label{appendix:homogeneity}
\input{homogeneity}

\section{Imaging beam}
\input{imaging_beam}

\section{Imaging fidelity}
\input{imaging_fidelity}

\section{Branching ratio}\label{appendix:branching_ratio}
\input{BranchingRatio}
\section{Temperature measurements}\label{appendix:release_recapture}
\input{release_recapture}
\section{Trapped atoms lifetime}\label{appendix:lifetime}
\input{Lifetime}

\end{document}

%% file: homogeneity.tex
We homogenize the traps by imaging the array after the AOD (AA Opto Electronic DTSXY-400-532-002) but before the microscope objective by reflecting a fraction of the incoming beam off of a beam sampler. 
The beam sampler's angle with respect to the propagation of the trapping light from the AOD to the microscope is minimized so as not to distort the image of the traps. 
We verified that the reflected intensities are proportional to the transmitted ones so that when the imaged intensities are homogeneous the transmitted ones are homogeneous as well. 
The RF signal used to create the array of traps is generated by an arbitrary waveform generator (Spectrum M4i.6621-x8 AWG) followed by an amplifier before being sent to the AOD. 
The AWG produces a set of sine waves at equally spaced frequencies. 
When all the tones are in phase the maximum voltage amplitude scales linearly with the number of traps and quickly saturates the amplifier. 
To avoid this we optimize the phases to minimize the signal's envelope, following the same protocol as in \cite{Endres2016}.
We finally image the trap intensities and feedback altering each of these tones' amplitude sequentially to minimize the trap intensities' variance. We finally obtain a standard deviation of trap intensities of about 2\,\%. When measuring the $\ket g \leftrightarrow \ket e$ transition frequency in non-magic conditions, we did not observe a significant inhomogeneity of the traps.

In our magic-polarization tweezers, the polarization homogeneity over all traps is important. We find that the acousto-optic deflector can lead to polarization inhomogeneity of the order of a degree or more for linear polarization. To prevent this polarization inhomogeneity, we placed a polarizer directly after the AOD. It turns a polarization inhomogeneity into power inhomogeneity, which is corrected by the feedback discussed above.
\change{
On the $5\times5$ arrays presented here, we find that this is enough to achieve a homogeneous polarization across the traps and we observed that the magic condition is satisfied simultaneously on all of them. If a slight inhomogeneity of polarization was observed, one might for instance compensate it by adjusting trap powers. 
}

%% file: imaging_beam.tex
\change{To image the atoms in the tweezers by fluorescence, we use a single non retro-reflected beam with components along both radial and axial directions of the tweezer.
We did not find that retro-reflecting the beam made any significant improvement in terms of cooling, and it caused light to be reflected on the glass cell and into the microscope objective which increased the background stray light. Similarly, adding beams along other directions tends to increase the amount of stray light reaching the camera and to decrease the image quality.
We assume that a single beam is enough to ensure efficient cooling because although the radiation pressure exerted by the beam is not balanced, the restoring force of the trap prevents the atom from being pushed away, and allows for continuous cooling.  }

%% file: imaging_fidelity.tex
To estimate the imaging fidelity $F$, we assume that the histogram of the number of collected photons follows a simple model:
If no atom is present in the trap, we collect on average $N_0$ photons due to the background light. The probability that $n$ photons reach the camera is then given by $P^P_{N_0}(n)$, where where 
$P^P_\lambda(n)=\lambda^n e^{-\lambda}/n!$
is the Poisson distribution with mean $\lambda$. 
In addition to the shot noise, the distribution is also broadened by the Gaussian camera readout noise and the probability to have a count $x$ is $\sum_{n=0}^{+\infty} P^P_{N_0}(n) g_{n, \sigma}(x)$ where $g_{n, \sigma}$ is a Gaussian distribution of mean $n$ and standard deviation $\sigma=1.6$. (This value is given by the read noise of the camera for a single pixel multiplied by the square root of the number of pixels over which the fluorescence is integrated.)

Similarly, if an atom is present throughout the total duration of the imaging, it scatters $N_1$ photons on average and the probability to have a count $x$ is $\sum_{n=0}^{+\infty} P^P_{N_0 + N_1}(n) g_{n, \sigma}(x)$.

The last possibility corresponds to an atom being lost during the imaging, after scattering a random number of photons $M$ between $0$ and $N_1$. The probability to collect $n$ photons is then
\begin{equation*}
    P^L_{N_0, N_1}(n) = \frac{1}{N_1}\int_{0}^{N_1}P^P_{N_0 + M} (n){\rm d}M
\end{equation*}
This corresponds to a smooth flat distribution that bridges the peaks for the zero and one atom cases.

\begin{figure}
    \centering
    \includegraphics{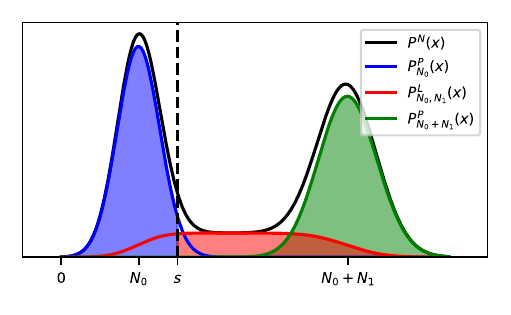}
    \caption{Distribution used to model the number of photons collected on the camera. The blue, green and red curves  correspond to the case where no atom is present in the trap, where the atom is initially present and survives the imaging and where an atom is lost during imaging, respectively. The black curve is the sum of all three distributions. The sum of the blue, green and red areas is the imaging fidelity for the threshold marked by the dashed line.}
    \label{fig:imaging_fidelity}
\end{figure}

Combining all three possibilities as shown on Fig.\,\ref{fig:imaging_fidelity}, the probability that $n$ photons reach the camera is 
\begin{multline*}
    P^N(n) = P_0 P^P_{N_0}(n) \\ + (1-P_0)\left[(1-P_{\rm loss}) P^P_{N_0 + N_1}(n) + P_{\rm loss} P^L_{N_0, N_1}(n) \right] 
\end{multline*}
where $P_0\simeq\SI{50}{\percent}$ is the probability that the trap is initially empty and $P_{\rm loss}$ is the probability to lose an atom while imaging it.

Taking into account the Gaussian noise of the camera, the probability to measure a count $x$ is then 
\begin{equation*}
    P^X(x) = \sum_{n=0}^{+\infty} P^N(n) g_{n, \sigma}(x)
\end{equation*}
The mean of this distribution is 
\begin{equation*}
\langle X \rangle = N_0 + N_1(1-P_0)(1-P_{\rm loss} /2 )
\end{equation*}
and its variance is 
\begin{multline*}
{\rm Var}(X) = \sigma ^ 2 + N_0
            + N_1 (1 - P_0)\times\\
            \left[
                N_1 \left(P_0  (1 - \frac{P_{\rm loss}}{2}) ^ 2 + \frac{P_{\rm loss}}{3} (1 - \frac{3 P_{\rm loss}}{4}  )\right)+ (1 - \frac{P_{\rm loss}}{2})\right]
\end{multline*}

This distribution is characterized by the set of parameters $P_0$, $P_{\rm loss}$, $N_0$, $N_1$ and $\sigma$ that we need to estimate to compute the fidelity as a function of the imaging parameters for example on figure \ref{fig:losses_vs_exposure}(a).

We record the histogram in the case were the tweezers are not loaded, which correspond to setting $P_0=1$. In this case $\langle X \rangle = N_0$ and ${\rm Var}(X) = N_0 + \sigma ^ 2$ so we can extract $N_0$ and $\sigma$.

To measure $P_{\rm loss}$ we record three pictures and measure the probability that the middle picture removes an atom. The last two parameters $P_0$ and $N_1$ can be extracted from the mean and variance of the distribution when the tweezers are loaded normally.\newline

Once the parameters have been estimated, one can compute the fidelity of the imaging $f(s)$ as a function of the threshold $s$ used to classify the presence of an atom.
The fidelity $f(s)$ is defined as the probability to correctly label the initial presence of an atom in the trap.
There are three events that contribute to the imaging fidelity~: no atom is present in the trap and the fluorescence collected is below the threshold; the atom is lost during imaging but scatters more photons than the threshold, or the atom is kept during the full imaging and scatter enough photons such that the fluorescence is higher than the threshold.
The probabilities of these events are respectively the blue, red and green areas on figure \ref{fig:imaging_fidelity}.
The sum of these three contributions gives us the imaging fidelity for a given threshold:
\begin{multline*}
    f(s) = P_0 P^P_{N_0}(X<s)  \\+ (1-P_0)\left[(1-P_{\rm loss}) P^P_{N_0 + N_1}(X>s) + P_{\rm loss} P^L_{N_0, N_1}(X>s) \right] 
\end{multline*}
We then compute the optimal fidelity $F$ and the best threshold $s_0$ by maximizing $f$. This method is used to calculate the fidelity in the rest of the text.

It is worth remarking that even if the zero and one atom peaks (blue and green curves on fig. \ref{fig:imaging_fidelity}) have negligible overlap, the imaging fidelity does not reach \SI{100}{\percent} because a fraction of the red curve is below the threshold.
This corresponds to the case where atoms are lost before having scattered enough photons to be distinguished from the background and this eventually limits our imaging fidelity to $F=\SI{99.1(0.2)}{\percent}$.

%% file: BranchingRatio.tex
Here we describe how we extracted the branching ratio of trapped to non-trapped metastable states: When imaging, we have measured a loss probability from $\ket g$ of about 6\,\% in \SI{30}{\milli\second}, which yields a loss rate of $\gamma_{\rm g} \simeq \SI{2}{\per\second}$. 
These atoms leak in two channels: towards trapped metastable states (which we call here $\ket t$) with a rate $\alpha\gamma_{\rm g}$, and towards non-trapped states with a rate $(1-\alpha)\gamma_{\rm g}$ with $\alpha$ the branching ratio. 
The $\ket t$ atoms can then decay back to $\ket g$, with a rate $\gamma_{\rm t-g}$. Under the application of imaging light, the atom numbers in $\ket g$ and $\ket t$ then follow the rate equations \begin{eqnarray*}
\dot n_{\rm g}=-\gamma_{\rm g}n_{\rm g}+n_{\rm t}\gamma_{\rm t-g}\\
\dot n_{\rm t}=\alpha\gamma_{\rm g}n_{\rm g}-n_{\rm t}\gamma_{\rm t-g}
\end{eqnarray*}

First, we extract $\gamma_{\rm t-g}$. For this we applied a \SI{1.5}{\second} imaging pulse and then removed the atoms remaining in $\ket g$. After this, the atoms that are in $\ket t$  will eventually decay back to $\ket g$ with a rate $\gamma_{\rm t-g}$. Fig.\,\ref{fig:metastable}(b) shows the fraction of atoms that have disappeared during the pulse, that re-appear in $\ket g$ as a function of wait time. By fitting the data with an exponential saturation, we extract a decay rate $\gamma_{\rm t-g}\approx \SI{2}{\per\second}$.  Such long lifetimes are similar to that observed with blue MOTs \cite{Youn2010}.
\begin{figure}
    \centering
    \includegraphics{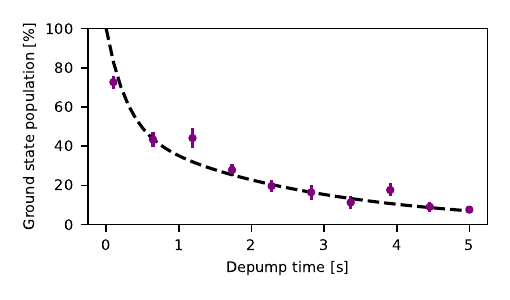}
    \caption{Measurement of the branching ratio $\alpha$ between the trapped and non-trapped metastable states. Fraction of atoms remaining in $\ket g$ under continuous imaging as a function of time. The dashed curve is the fit for the ground state population given by the rate equations (see text) with $\alpha=0.65$.}
    \label{fig:branching_ratio}
\end{figure}

Finally, in Fig.\,\ref{fig:branching_ratio} we apply an imaging pulse of variable time, and plot the fraction of atoms remaining in $\ket g$ after the pulse. Fitting then the data of Fig.\,\ref{fig:branching_ratio} with the rate equations using the measured rates $\gamma_{\rm g}$, $\gamma_{\rm t-g}$ and leaving $\alpha$ as a free parameter, we obtain a good agreement with the data for $\alpha = 0.65$, (dashed line in Fig.\,\ref{fig:branching_ratio}).

%% file: release_recapture.tex
\begin{figure}
    \centering
    \includegraphics{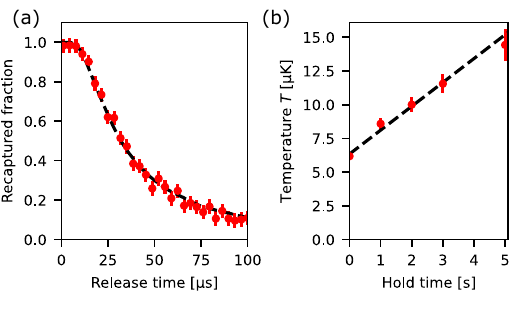}
    \caption{(a)~Measurement of temperature by release and recapture. The measured recaptured probability is corrected to account for imaging losses. The dashed line is a result of simulations using a waist of \SI{500}{\nano\meter}. The fit gives us a temperature of \SI{6.3}{\micro\kelvin}. (b)~Heating in the tweezers in absence of cooling light. The dashed line is a fit by a linear function to estimate the heating rate.}
    \label{fig:temperature}
\end{figure}

To measure the temperature of the atoms in the tweezers, we use the release and recapture technique \cite{Tuchendler2008}.
We suddenly turn off the tweezers for a few tens of microseconds and then switch them back on.
The fraction of recaptured atoms depends on the initial temperature of the atoms -- the lower the more likely for atoms to be trapped again -- and we extract the temperature by comparing with numerical simulations.
A typical temperature measurement is shown in \ref{fig:temperature}(a).

Just after the imaging step described in the main text, the temperature of the atoms is $T_0=\SI{6.3(0.2)}{\micro\kelvin}$.
In the absence of cooling light, the atoms slowly heat up in the tweezers. 
For a tweezer power $P_{\rm trap} = \SI{2.1}{\milli\watt}$ (trap depth of \SI{150}{\micro\kelvin}), we measure a heating rate of $\SI{1.7(0.2)}{\micro\kelvin\per\second}$ (see figure \ref{fig:temperature}(b)). This heating rate is compatible with expectations from the imaginary part of the polarizability \cite{Li2017}. We note that it is dominated by contributions from the broad transitions near \SI{400}{\nano\meter} rather than by the close narrow transition at \SI{530.3}{\nano\meter}. 

%% file: Lifetime.tex
\begin{figure}
    \centering
    \includegraphics{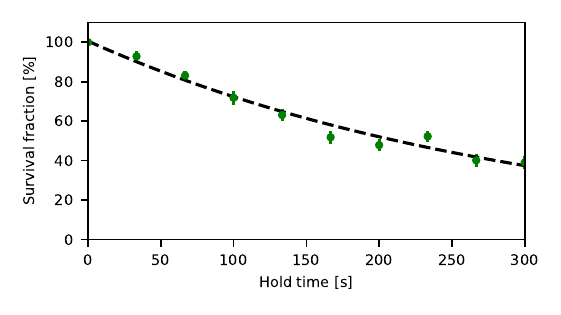}
    \caption{Fraction of atoms remaining in the tweezers as a function of hold time under continuous cooling. The measured fraction is corrected to account for imaging losses.}
    \label{fig:atom_lifetime}
\end{figure}

Fig.\,\ref{fig:atom_lifetime} 
shows the fraction of remaining atoms after a given hold time, under continuous cooling at $\Delta = -1.3\,\Gamma$, $I=10^{-3}\,I_{\rm sat}$, from which we extract an exponential lifetime of $\tau=\SI{300(30)}{\second}$, dashed line. This lifetime is due both to cooling-induced losses and vacuum : $\tau^{-1}=\tau_{\rm cool}^{-1}+\tau_{\rm vac}^{-1}$. By measuring lifetimes at higher cooling powers where vacuum loss rate is negligible, we extrapolate a linear dependence of the loss rate versus cooling power. This predicts that the lifetime due to cooling at $I=10^{-3}\,I_{\rm sat}$ of $\tau_{\rm cool}= \SI{350(100)}{\second}$. Comparing this expectation to the observation of $\tau=\SI{300(30)}{\second}$, we obtain a safe lower bound for the vacuum lifetime $\tau_{\rm vac}\geqslant\SI{500}{\second}$. 
\newpage